\documentclass[12pt]{article}

\usepackage[latin1]{inputenc}
\usepackage{graphicx}

\newcommand{\be}{\begin{equation}}
\newcommand{\ee}{\end{equation}}
\def\ba{\begin{eqnarray}}
\def\ea{\end{eqnarray}}

\title{
Fluctuating Initial Conditions and Anisotropic Flows }

\author{
Fernando G. \textsc{Gardim}$^{1,}$\footnote{E-mail: gardim@fma.if.usp.br}, Yogiro \textsc{Hama}$^{1}$
and Fr\'ed\'erique \textsc{Grassi}$^{1}$\\
\it{\small{$^1$Instituto de F\'\i sica, Universidade de S\~ao
Paulo, C.P. 66318, 05315-970,}}\\ \it{\small{S\~ao Paulo-SP, Brazil}}\\
}

\begin{document}
\maketitle

\abstract{
In this work we study the connection between anisotropic flows
and lumpy initial conditions for Au+Au collisions at 200GeV. We
present comparisons between anisotropic flow coefficients and
eccentricities up to sixth order, and between initial condition reference
angles and azimuthal particle distribution angles. We also present
a toy model to justify the lack of connection between flow
coefficients and eccentricities for individual events.

}

\section{Introduction}

The anisotropy in the azimuthal particle distribution in
relativistic heavy-ion collisions has been interpreted as an
indication of the creation of a strongly interacting Quark-Gluon
Plasma (QGP) in the liquid phase with low viscosity. However, this
anisotropy should reflect the initial spatial deformation of the matter
created. Recently, several analyses have been performed in an effort to understand the relation between anisotropic flows and the initial-conditions (IC) geometry \cite{Qiu:2011iv,Qin:2010pf}, pointing out a close relation
between the elliptic flow $v_2$ and the ellipticity of the IC
$\varepsilon_2$, and also between triangular flow $v_3$ and the
triangularity of the IC $\varepsilon_3$. Nevertheless, in all
these studies fairly well-behaved (smooth) initial condition were used.

In this article we present a comparison between initial
conditions and the anisotro-pic flow for very irregular (lumpy) initial
conditions, for Au-Au collisions at 200 GeV. We also present a toy
model, based on the One-Tube Model \cite{Hama:2009vu}, in order to
show why the eccentricity coefficients do not carry all the
necessary information to understand the final momentum anisotropy.

\section{Connection Between Flow and IC}\label{sec:1}

In order to compare IC anisotropies with the azimuthal
distribution of final-state hadrons we use the NeXSPheRIO code. This code is
based on event-by-event 3+1D hydrodynamics, where the IC are
generated by NeXuS, and SPheRIO solves the equations of
relativistic ideal hydrodynamics. The NeXSPheRIO code provides a
good description for several experimental data trends, for instance,
$v_2$ dependency on $\eta$ and $p_T$, and $v_2$ fluctuations
\cite{Hama:2007dq}, directed flow $v_1$ \cite{Gardim:2011qn}, and
reproduces the structure observed in two-particle correlations
\cite{Takahashi:2009na}.

The analysis presented in this work was done for 1000 events
equally divided into six centrality bins, and the hydrodynamics was
computed for each event. Charged particles at mid-rapidity are
used for the comparison. The anisotropic flow coefficients and
reference angles come from the Fourier expansion of the azimuthal
distribution of particles

\begin{eqnarray}
v_n e^{in\Psi_n}\equiv \langle e^{in\phi}\rangle, \label{vn}
\end{eqnarray}
where $\phi$ is the azimuthal momentum angle. The eccentricity coefficients come from

\begin{eqnarray}
\varepsilon_n e^{in\Phi_n}\equiv - \frac{\langle r^n e^{in\phi_s}
\rangle}{\langle r^n \rangle}, \label{en}
\end{eqnarray}
$\phi_s$ as spatial angle in the x-y plane. $\varepsilon_n$ is computed in the C.M. system.

\begin{figure}[h]
       \centerline{\includegraphics[width=11cm]
                                   {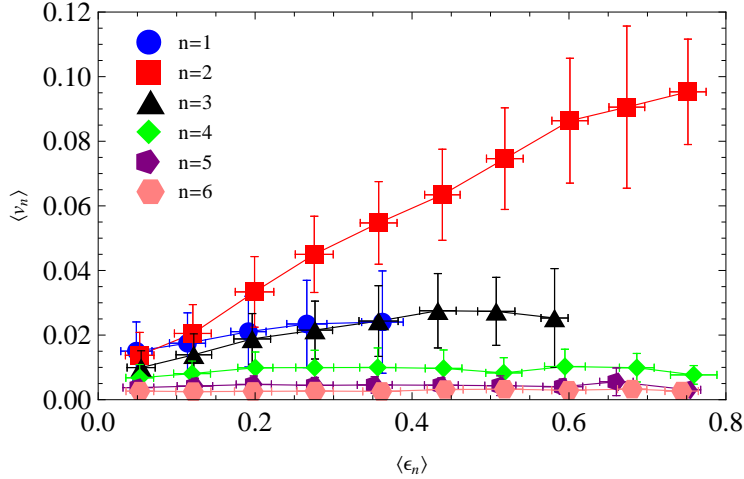}}
   \caption{Comparison between $v_n$ and $\varepsilon_n$ for charged particles at mid-rapidity for Au-Au at 200 GeV.}
   \label{fig:1}
\end{figure}

The comparison between average $v_n$ and $\varepsilon_n$, as well
as their fluctuations is shown in Fig. \ref{fig:1}. For $n\ge 4$ the
response to the average global initial geometry deformation, given
by $\langle\varepsilon_n\rangle$, is not completely reflected
in the flow coefficients. The average elliptic flow $\langle
v_2\rangle$ is proportional to the average ellipticity
$\langle\varepsilon_2\rangle$, except for $\varepsilon_2\ge 0.6$,
where the curve slope has a small deviation. As for $\langle v_1\rangle$, it slightly increases with $\langle\varepsilon_1\rangle$. The ratio $\langle v_1\rangle/\langle\varepsilon_1 \rangle$ is not constant as already
shown in Ref. \cite{Gardim:2011qn}. For the average triangular
flow, whose non-zero value is connected with the ridge effect
\cite{Alver:2010gr}, it is possible to observe two behaviors: for
$\langle\varepsilon_3 \rangle<0.4$ there is a direct relation
between triangularity and triangular flow, positive and constant
slope, however for $\langle\varepsilon_3 \rangle\ge 0.4$, $\langle v_3\rangle$
decreases. In comparison to Refs. \cite{Qiu:2011iv,Qin:2010pf}, we
have qualitatively the same result for $n=2$, but somewhat
different result for $n=3$. Our $\langle v_3
\rangle/\langle\varepsilon_3 \rangle$ is not constant.
Nevertheless an important result, in addition to the average
values comparison, is the large dispersion values. Note that for
lumpy IC, the eccentricity coefficients do not have all the
information to understand the anisotropic flow. Since they give
information about the global geometry, leaving aside the lumpiness
of the IC.

\begin{figure}[h]
       \centerline{\includegraphics[width=9cm]
                                   {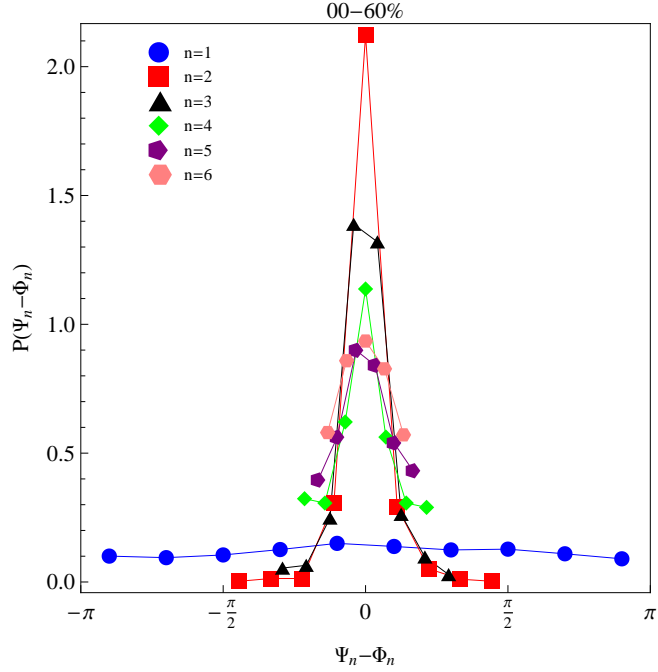}}
   \caption{Comparison between $\Psi_n$ and $\Phi_n$ for charged particles at mid-rapidity for Au-Au at 200 GeV. In abscissa axis is represented
   the difference $\Psi_n-\Phi_n$ and in the ordinate axis is plotted the probability density to get $\Psi_n-\Phi_n$. }
   \label{fig:2}
\end{figure}

Besides the connection between $v_n$ and $\varepsilon_n$, it is
important to compare the reference angles for IC $\Phi_n$ and flow
$\Psi_n$, since it is expected that their mean values are equals.
This comparison is plotted in Fig. \ref{fig:2}. Due to the almond collisional shape
of the IC, $\Psi_2$ is well correlated with $\Phi_2$. For $n=3$,
there is a correlation between the reference angle, but not as
strong as for $n=2$. However, for the other harmonics, the
dispersions are larger.

\section{Tube model for $\varepsilon_3$=0}

The NeXSPheRIO results presented in Sec. \ref{sec:1} show that
there is no strict event-by-event connection between $v_n$ and
$\varepsilon_n$. However, it is important to understand what
features of the lumpy IC are manifested in final-state anisotropic flows.
Inspired by the One-Tube Model \cite{Hama:2009vu}, we create an IC
with triangularity approximately equal to zero, and observe what
happens with the triangular flow. The One-Tube Model provides an
understanding of the ridge origin, and can be summarized as
follows: NeXus IC conditions have tubular structures (along the
collision axis). Using an IC composed of a smooth background
energy density in the transverse plane plus one typical tube (for
central collisions), we obtain two correlated peaks in the
azimuthal distribution of particles. Due to this correlation, the
2-particle correlation has the 3 peak structure seen in the data
\cite{Putschke:2007if}. However, for this correlation to appear,
it is necessary that the tube be positioned near the boundary.

\begin{figure}[h]
\includegraphics[width=5cm]{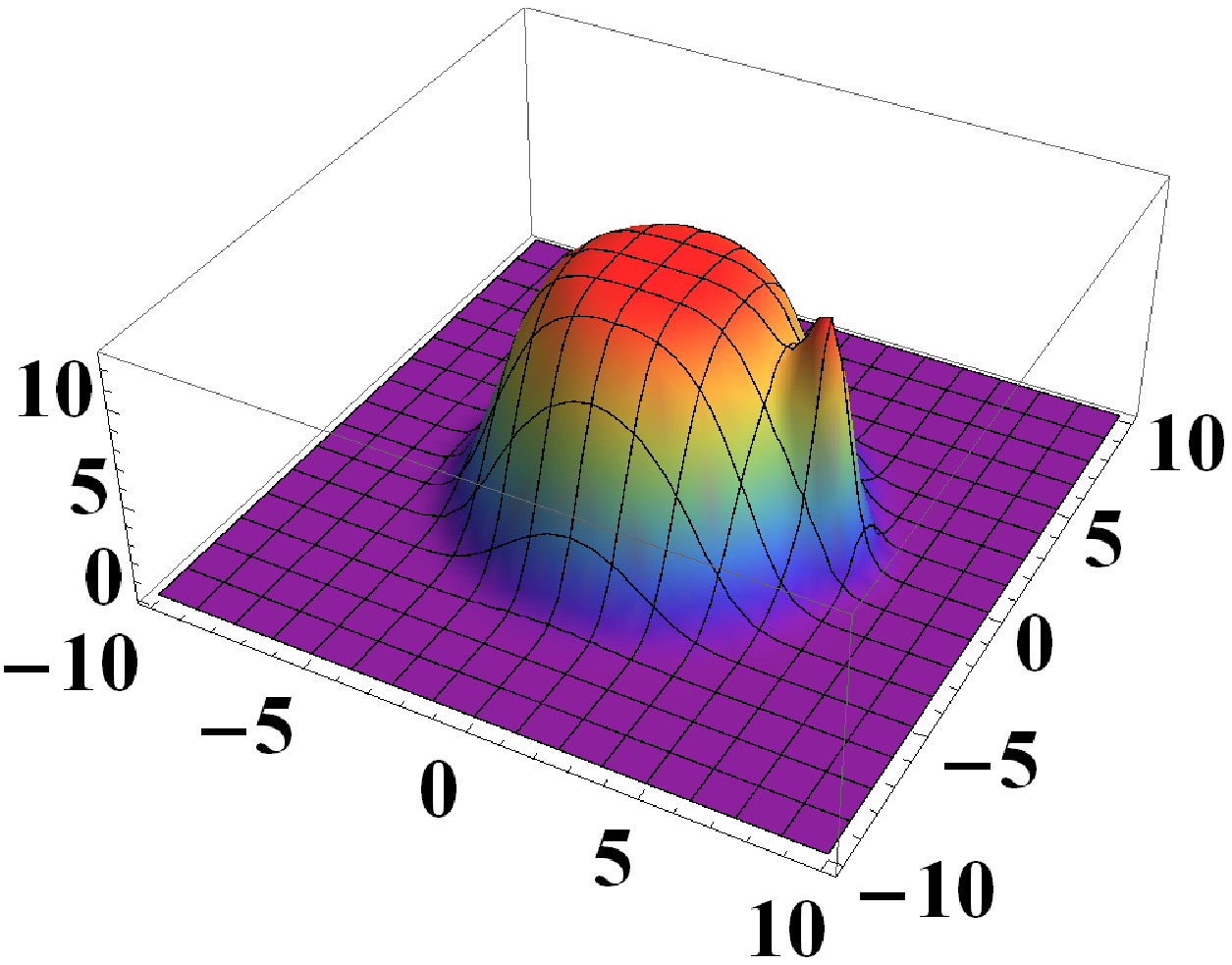}
\includegraphics[width=5cm]{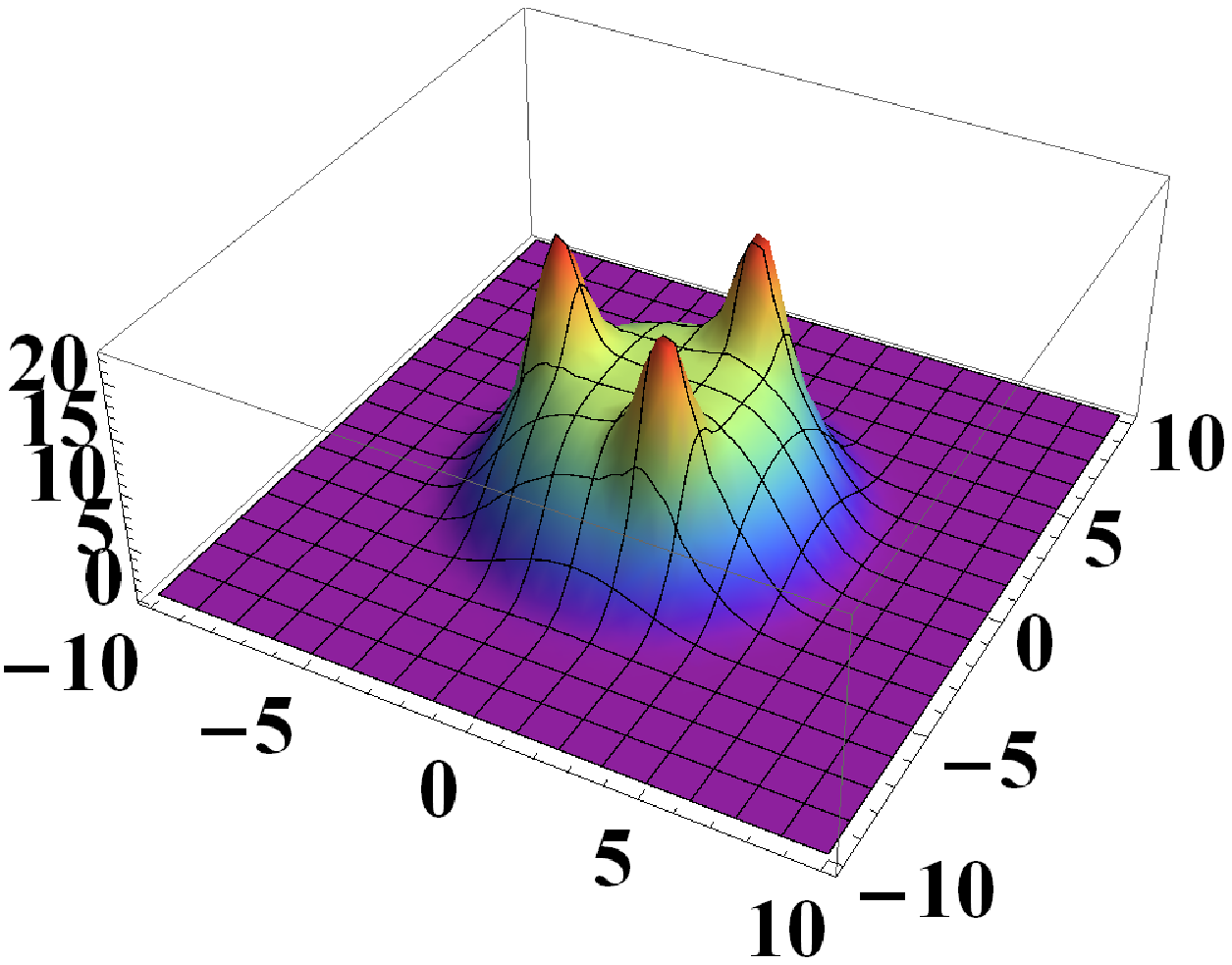}
\includegraphics[width=5cm]{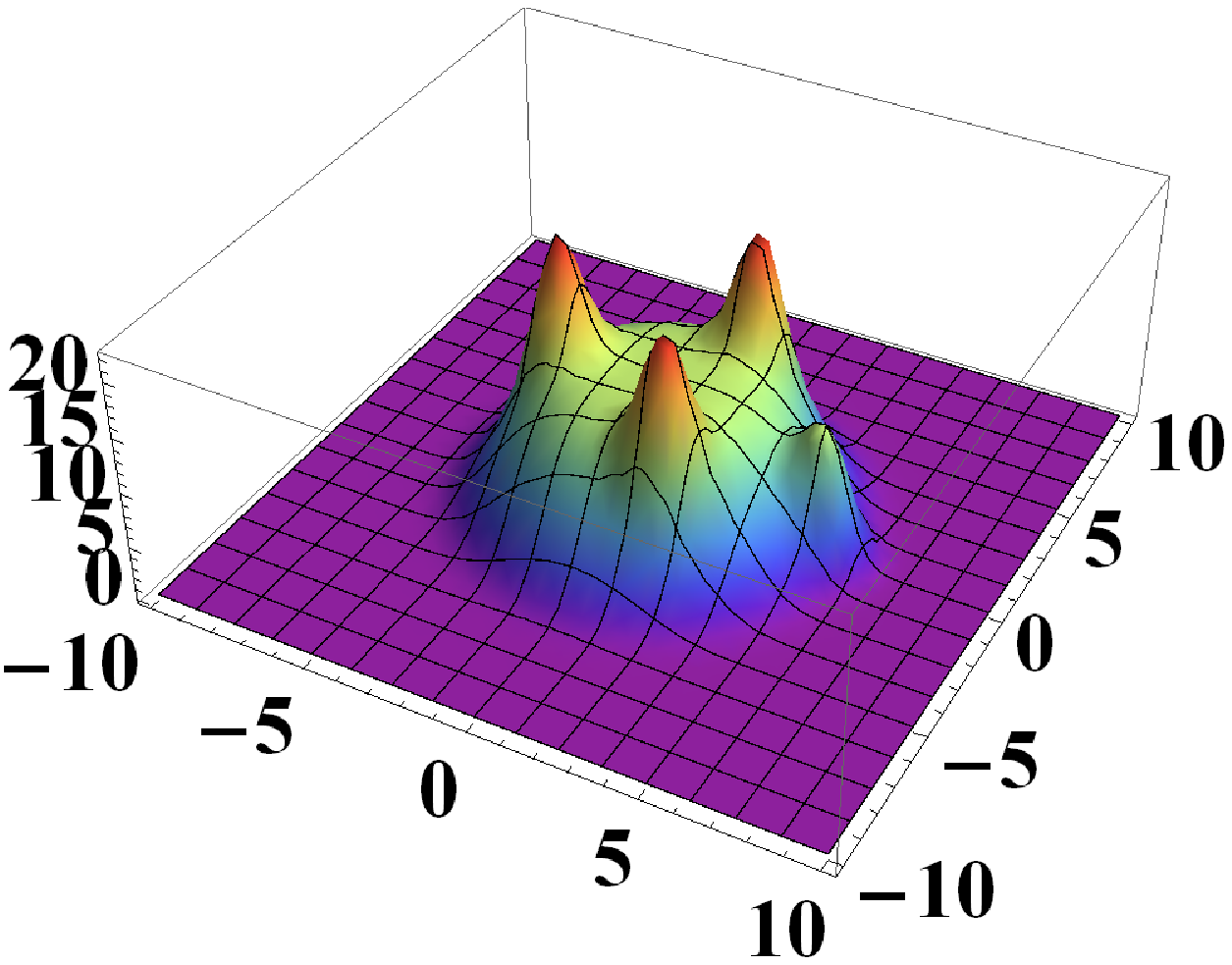}

\caption{Energy density profiles for:(Left) One-Tube model, (center) 3 inner tubes, and (right) IC for $\varepsilon_3=0$.}
\label{fig:3}
\end{figure}

Keeping in mind the One-Tube model for central collisions, we
compare three different IC. The first one is the One-Tube model,
left plot in Fig. \ref{fig:3}, with the triangularity
$\varepsilon_3^{(1t)}\simeq 0.067$. Then, we create an IC
condition with three equal inner tubes, but with the same
triangularity $\varepsilon_3^{(3t)}\simeq 0.067$, center plot in
Fig. \ref{fig:3}. Finally, we consider an IC without triangularity
$\varepsilon_3^{(4t)}\simeq 0$, right plot Fig. \ref{fig:3}. For
these IC, we use a 2+1D version of the SPheRIO code, with
longitudinal boost invariancy, to solve the ideal hydrodynamics.

The triangular flow obtained was: for the three inner tubes
$v_3^{(3t)}\simeq 0.00095$,  showing that the lumpiness in the
internal region makes only a small contribution to the final anisotropy.
However, for the ICs with the same peripheral tube, but with
different inner region, almost the same value of the triangular
flow was obtained: $v_3^{(1t)}\simeq 0.0110$ and $v_3^{(4t)}\simeq
0.0116$, respectively for the One-Tube case and $\varepsilon_3=0$
case.

If one compares all the flow harmonics, it can be seen that for
the IC with 3 inner tubes, not only $v_3$ is small, but all flow
harmonics are closer to zero, leading to an almost isotropic
distribution. For the other two ICs we conclude that the flow
coefficients are almost the same, even though with different
eccentricity coefficients. This result stresses the fact that
there is no strict event-by-event relation between flow and
eccentricity.

\section{Conclusions}

Using lumpy initial conditions for Au+Au at $\sqrt{s}=200 AGeV$ and
solving hydrodynamics event-by-event, it was shown that $\langle
v_n\rangle$ is not proportional to $\langle\varepsilon_n\rangle$,
except for the $n=2$ case, which is dominated by geometry effects.
In event-by-event hydrodynamics, there is no direct relation
between $v_n$ and $\varepsilon_n$, since the lumpiness is not
completly captured by the eccentricity definition given by Eq.
\ref{en}. Using NeXus and the Tube-Model IC, we understand that
peripheral matter is more important than the inner matter for the
momentum anisotropy in the final state at central collisions. This
same behavior happens for peripheral collisions, but we cannot
neglect the elliptic shape of initial overlap geometry as a significant contribution to anisotropic flow.


\section*{Acknowledgements}
This work is funded by FAPESP under projects 09/16860-3 and 09/50180-0, and CNPq.

\end{document}